\documentclass[CEJP,PDF]{cej} 
\usepackage{layout}
\usepackage{amsmath}
\usepackage{amssymb}
\usepackage{amsthm}
\usepackage{textcomp}
\usepackage{hyperref}

\title{Self-oscillation acoustic system destined to measurement of stresses in mass rocks}

\articletype{Editorial}

\author{Janusz Kwa\'{s}niewski\inst{1}\email{kwa\_j@agh.edu.pl},
       Yury  Kravtsov\inst{2}\email{y.kravtsov@am.szczecin.pl},
      Ireneusz Dominik\inst{2}\email{dominik@agh.edu.pl}
      Lech Dorobczynski\inst{1}\email{l.dorobczynski@am.szczecin.pl}}

\institute{
     \inst{1} Faculty of Mechanical Engineering and Robotics \\AGH University of Science and Technology\\
     Al. Mickiewicza 30, 30-059 Krak\'{o}w , Poland
     \inst{2} Faculty of Marine Engineering\\Maritime University of Szczecin\\
     Ul. Wa\l{}y Chrobrego 1-2, 70-500 Szczecin, Poland
          }

\abstract{The paper presents an electronic self-oscillation acoustic system (SAS) destined to measure of stresses variations in the elastic media. The system consists of piezoelectric detector, amplifier-limiter, pass-band filter, piezoelectric exciter and the frequency meter. The mass rock plays a role of delaying element, in which variations in stresses causing the variations of acoustic wave velocity of propagation, and successive variation in frequency of oscillations generated by system.\\The laboratory test permitted to estimate variations in frequency caused by variations in stresses of elastic medium. The principles of selection of frequency and other parameters of the electronic system in application to stresses measurement in condition of the mine were presented.}


\begin{document}
\maketitle


\section{Introduction}

Autodyne systems are widely applied in radio engineering for monitoring of changes in tested object [5,6]. In mechanics and geophysics autodyne systems are used much rarely.  Given paper describes a self-exiting acoustic system of autodyne type destined to measurements of stressed changes in elastic media.\\The structure of mentioned systems consists of amplifier with feedback loop of such properties, that the entirely system is unstable and generates oscillations.  The function of sensitive element is performed by elements of feedback loop, which properties are depending on variations of measured value. As the final effect the frequency of self-oscillations is a measure of variations in input value. 

\section{Basic equations of SAS}

\begin{figure}[!ht]
\includegraphics[width=12cm]{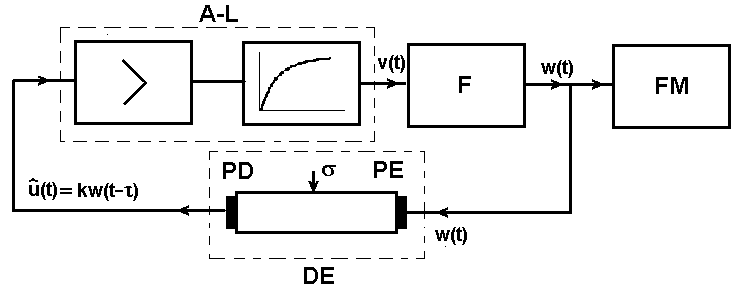}
\caption{Basic  schema of self-oscillating acoustic system which varies  frequency under influence of stress in delaying element}
\label{fig:f1}
\end{figure}

Let’s consider self-oscillation acoustic system, presented in Fig. \ref{fig:f1}, which consists of 

\begin{itemize}
\item piezoelectric detector PD
\item amplifier-limiter A-L
\item pass-band filter F
\item piezoelectric exciter  PE 
\item delaying element DE 
\item frequency meter FM
\end{itemize}

Input of A-L is a harmonic signal described by amplitude $U$, frequency $\omega$ and phase angles $\varphi_u$:

 \begin{equation}
\label{eq:n1}
 u(t)=U\cos(\omega t + \varphi_u)
\end{equation}

Output  signal $v(t)$  of non-linear A-L contains the basic frequency as well higher harmonics:

\begin{equation}
 v(t)=F_{\omega}(v)\cos(\omega t+\varphi_1)+F_{2\omega}(v)\cos(2\omega t + \varphi_2)+...
\label{eq:e2}
\end{equation}

Amplitudes of harmonic components are depending on amplitude of input signal U. The most important is the first component of sum (\ref{eq:e2}):

\begin{equation}
 v(t)=F_{\omega}(v)\cos(\omega t+\varphi_1)
\label{eq:e2}
\end{equation}

\begin{figure}[!ht]
\includegraphics[width=10cm]{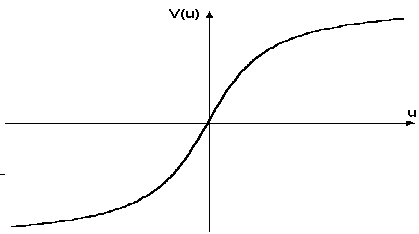}
\caption{Non-linear dependence between output amplitude of A-L and input amplitude $U$}
\label{fig:f2}
\end{figure}

Non-linear characteristic of amplifier-limiter is shown in Fig. \ref{fig:f2}, under assumption, that an amplifier is saturated after increasing of amplitude $U$.

\begin{figure}[!ht]
\includegraphics[width=10cm]{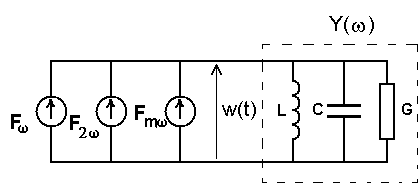}
\caption{Filtration of higher harmonics in the resonant circuit}
\label{fig:f3}
\end{figure}

Due to elimination of higher harmonic, signal v is transmitted by pass-band filter F, realised as parallel connection of inductivity L, capacity  C and conductance G , which is shown in Fig. \ref{fig:f3}

Admittance of such filter for the frequency $\omega$ can be written as: 

\begin{equation}
Y(\omega)=i\omega C + \frac{1}{i\omega L}+ G,
\end{equation}

where the real part of the admittance G contains the inner conductance of current source $G_z$, component $G_F$ describes  energy loss in the filter, and finally $G_w$ component describes energy loss in piezoelectric exciter:

\begin{equation}
G=G_Z+G_F+G_w.
\end{equation}

The output signal of A-L was presented as parallel connection of current sources, corresponding with successive harmonics $\omega,2\omega,...m\omega$. The basic component of $W(t)$ on the filter output can be described as 

\begin{equation}
\label{eq:e0}
W(t)=W_\omega\cos\left[\omega t+\varphi_w(\omega)\right]
\end{equation}

where $W_\omega$ denotes the absolute value, and $\varphi_w$ - phase angle of the complex amplitude: 

\begin{equation}
\hat W_\omega = W_\omega e^{i\varphi_w}
\end{equation}

which is connected with amplitude of first harmonic $F_1(U)$ by relationship:

\begin{equation}
\label{eq:e3}
\hat W_\omega = \frac{F_1(U)e^{i\varphi_u}}{Y(\omega)},
\end{equation}

where $Y(\omega)$ denotes filter admittance. It implies that:

\begin{equation}
W_\omega = \frac{F_1(U)e^{i\varphi_u}}{|Y(\omega)|}
\end{equation}

as well as :

\begin{equation}
\varphi_w=\varphi_u-\arg\left[ Y(\omega)\right]=\varphi_u-\arctan\left(\frac{\omega C - \frac{1}{\omega L}}{G}\right)
\end{equation}

The complex amplitudes of higher harmonics are described using formulae analogous to (\ref{eq:e3})

\begin{equation}
\label{eq:e11}
\hat W_{m\omega} = \frac{F_m(U)e^{i\varphi_u}}{Y(m\omega)},
\end{equation}

The absolute value of admittance $Y(m\omega)$ is  significantly greater than absolute value of  admittance $Y(\omega)$ for values of $\omega$, nearly to resonant frequency $\omega_0=1/\sqrt{LC}$, for which $Y=(\omega)\approx G$, so that : $|\hat W_{m\omega}| \ll |\hat W_\omega|$,  which permits to omit the influence  of higher harmonics, and to perform the analysis using only  component (\ref{eq:e0}).

Voltage signal (\ref{eq:e0}) drives the input of PE, exciting acoustic waves in delaying element DE, forming the feedback voltage:

\begin{equation}
\label{eq:e12}
\hat u(t)=kw(t-\tau)=k W_\omega \cos(\omega t - \omega\tau+\varphi_w)
\end{equation}

where:\\
$\tau$ - delay time,\\
k-coefficient of damping and energy lossess in delaying element.

Substituting (\ref{eq:e11}) to (\ref{eq:e12}) we obtain:

\begin{equation}
\label{eq:e13}
\hat u(t)=k\frac{F_\omega(U)}{|Y(\omega)|}\cos(\omega t - \omega\tau+\varphi_w-\arg Y(\omega))
\end{equation}

Comparing (\ref{eq:e13}) with the formula describing input signal of the amplifier (\ref{eq:n1}), we can to obtain the equation of phase balance: 

\begin{equation}
\label{eq:e14}
\omega\tau+\arg Y(\omega)=2\pi n
\end{equation}

which determines the frequency of self-oscillations. Simultaneously we have the following equation of amplitude balance,

\begin{equation}
U=k\frac{F_\omega(U)}{|Y(\omega)|}
\end{equation}

which permits to estimate the necessary gain factor of A-L element and to determine the amplitude of self-oscillations.

\section{Relationship between frequency of self-oscillation and stresses}

\begin{figure}[!ht]
\includegraphics[width=8cm]{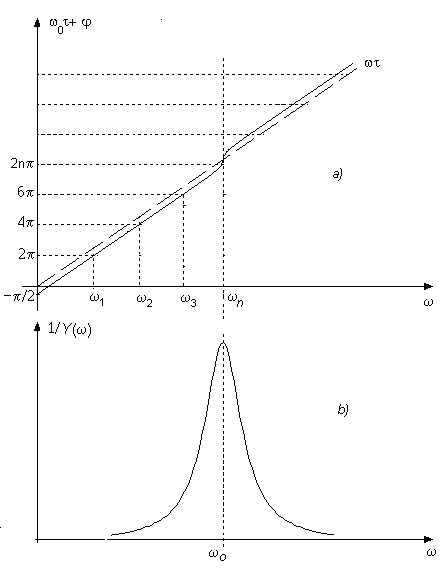}
\caption{a) Principle of self-oscillations generation, b) Filter frequency response}
\label{fig:f4}
\end{figure}

The dependence of left side of phase balance equation (\ref{eq:e14}) on the frequency is  shown in Fig. \ref{fig:f4}b, where:   

\begin{equation}
\varphi(\omega)=\arg Y(\omega) =\arctan\left[(\omega C - \frac{1}{\omega L})/G\right].
\end{equation}

The pass band of resonant circuit of Q factor $\Delta\omega=\left(\omega_0-\frac{\omega_0}{2Q},\omega_0+\frac{\omega_0}{2Q}\right)$ can cover the some frequencies $\omega_n$ corresponding with the different values of index n. To avoid generation of multiple frequencies, the following inequality has to be fulfilled:

\begin{equation}
\omega_{n+1}-\omega_{n} > \frac{\omega_0}{Q}.
\end{equation}

Using the estimation :

\begin{equation}
\omega_n\approx \frac{2\pi n}{\tau}
\end{equation}

which corresponds the omitting of component in (\ref{eq:e14}) we have:

\begin{equation}
\frac{1}{\tau} > \frac{\omega_0}{Q} \Rightarrow Q > \omega_0\tau
\end{equation}

In such conditions frequency can be determined as follows: 

\begin{equation}
\label{eq:e20}
\omega\tau=2\pi n .
\end{equation}

Deriving (\ref{eq:e20}) with stress $\sigma$ as the independent variable we obtain:

\begin{equation}
\label{eq:e21}
\frac{\operatorname{d}\omega}{\operatorname{d}\sigma}+\omega\frac{\operatorname{d}\tau}{\operatorname{d}\sigma}=0 \Rightarrow \frac{1}{\omega}\frac{\operatorname{d}\omega}{\operatorname{d}\sigma}\tau=-\frac{1}{\tau}\frac{\operatorname{d}\tau}{\operatorname{d}\sigma}
\end{equation}

Formula (\ref{eq:e21}) is a basic to determine the stresses variation using SAS. Let’s take in account phase component  in equation (\ref{eq:e14}) by approximation:

\begin{equation}
\label{eq:e22}
\varphi(\omega)\approx\frac{\operatorname{d}\varphi(\omega_0)}{\operatorname{d}\omega}(\omega-\omega_0)=\frac{2Q}{\omega_0}(\omega-\omega_0)\equiv T(\omega-\omega_0)
\end{equation}

Where $T=\frac{2Q}{\omega_0}=\frac{T\omega}{\pi}$ denotes time constant of resonant circuit.

Substituting (\ref{eq:e21}) into (\ref{eq:e13}) we have:

\begin{equation}
\label{eq:e23}
\omega T+T(\omega-\omega_0)=2\pi n
\end{equation}

and:

\begin{equation}
\label{eq:e24}
\frac{\operatorname{d}\omega}{\operatorname{d}\sigma}(\tau+T)+\omega\frac{\operatorname{d}\tau}{\operatorname{d}\sigma}=0
\end{equation}

As result, instead equation (\ref{eq:e21}) we have equation:

\begin{equation}
\label{eq:e25}
\frac{1}{\omega}\frac{\operatorname{d}\omega}{\operatorname{d}\sigma}=-\frac{1}{\tau+T}\frac{\operatorname{d}\tau}{\operatorname{d}\sigma}
\end{equation}

where $\frac{1}{\tau}$ component was changed into $\frac{1}{\tau+T}$. Usually $ T \ll \tau$ and for this reason derivative $\frac{\operatorname{d}\omega}{\operatorname{d}\sigma}$ in (\ref{eq:e25}) is almost equal to (\ref{eq:e21}).

Equation (\ref{eq:e23}) makes possible to estimate the sensitivity of variation in frequency caused by variation in temperature $t^0$:

$$\frac{\operatorname{d}\omega}{\operatorname{d}t^0}(\tau+T)-T\frac{\operatorname{d}\omega_0}{\operatorname{d}t^0}$$

which allows to write: 

\begin{equation}
\label{eq:e26}
\frac{\operatorname{d}\omega}{\operatorname{d}t^0}=\frac{T}{\tau+T}\frac{\operatorname{d}\omega_0}{\operatorname{d}t^0}
\end{equation}

As result we obtained that, the sensitivity of variation in frequency caused by variation in temperature is $\frac{T}{\tau+T}$  times lower than the sensitivity of variation in resonant frequency caused by variation in temperature.

\section{Estimation of SAS sensitivity}

\begin{figure}[!ht]
\includegraphics[width=8cm]{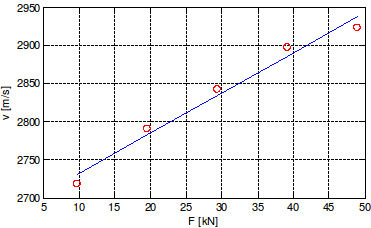}
\caption{Dependency between velocity of sound  and pressing force in concrete}
\label{fig:f5}
\end{figure}

Computer simulations performed  in MATLAB-Simulink environment were based on preliminary research performed in laboratories of AGH University of Science and Technology in Krakow. The mentioned research was concerned on determination of dependency between compressing stress and velocity of sound in the specimen made of concrete. Obtained linear regressive model has a form: $v = a \cdot F+b$\\
			
 where:\\
F - compressing force expressed in  kN, \\
v - sound velocity expressed  in m/s. \\
Values of coefficients a and b are as follows: a = 5.27807339449537  [m/Ns],   b = 2678.4917 \\
Obtained results are presented in Fig \ref{fig:f5}. 

Simulation was performed for specimen length L=0.45 m.
	Parameters of model built in Simulink were as below:
\begin{itemize}
\item static characteristic of limiter was described by formula $y=\arctan(x)$.
\item gain factor of amplifier was assumed as K=25
\item band pass filter, consisted of two sections, each of them described by transfer function:
\end{itemize}

\begin{equation}
\label{eq:e27}
G_F(s)=\frac{\omega_m s}{Qs^2+\omega_m s +Q \omega_m^2}
\end{equation}

where:\\
$\omega_m=\sqrt{\omega_1\omega_2}$,\\
$Q=\frac{\omega_m}{\omega_2-\omega_1}$,\\
$\omega_1=30787.61 \frac{\operatorname{rad}}{s}$, $\omega_2=30787.61 \frac{\operatorname{rad}}{s}$ - are lower and upper cut-off frequencies of the filter.
Obtained resonant frequencies in dependence of pressing force are presented in Table \ref{tab:t1}

\begin{table}[!ht]
\label{tab:t1} 
\centering
\begin{tabular}{ | c | c | c | c | c | c | }
\hline
  $F$ [kN] & 9.81   & 19.62 & 29.43 & 39.42 & 49.05 \\ \hline
  $v$ [m/s]  & 2717 & 2768 & 2819 & 2871&2922 \\ \hline
  $f_0$ [Hz]  & 4666 & 4676 & 4688 & 4694 & 4702 \\ \hline

\end{tabular} 
\end{table}

Basing on previous data, system sensitivity factor, expressed in form:

\begin{displaymath}
\label{eq:e27}
S=\frac{\Delta f_0}{\Delta F}
\end{displaymath}

can be estimated as:

\begin{equation}
\label{eq:e27}
S=1.02 \frac{\operatorname{Hz}}{\operatorname{kN}}
\end{equation}

\section{Laboratory stand}

\begin{figure}[!ht]
\includegraphics[width=13cm]{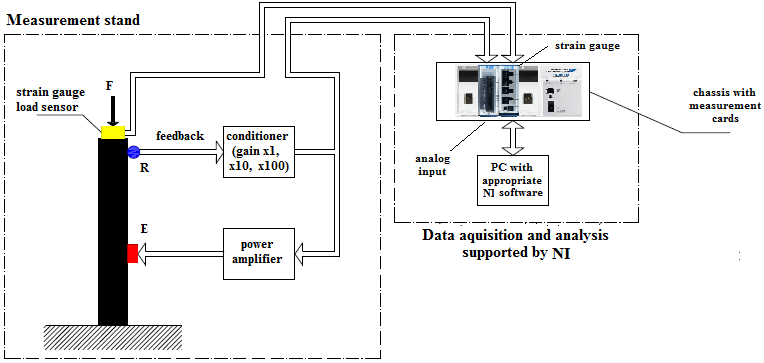}
\caption{Self-exited Acoustical System diagram, where E - shaker and R - receiver}
\label{fig:f6}
\end{figure}

During our preliminary research at the Department of Process Control at the AGH University of Science and Technology the Self-exited Acoustical System has been developed. System diagram is shown in figure \ref{fig:f6}, where amplifier, shaker (E) and receiver (R) are formed in a feedback loop. The shaker (E) was fixed to a stone bar. The bar was put into a testing machine to create a load. On the beam’s surface four accelerometers were fixed: three of them on the same surface as the shaker and the last one directly on the opposite site.

As a result of positive feedback, there is a bilateral interaction between the control device and the vibrating system, which allows the self-oscillating system to control its own energy balance. As a result of that, despite the fact of losses of energy occurring in the system, there are non-fading periodic oscillations. 
Many applications of the self-oscillating phenomenon are known. These are: the vibrations of cutting tools, turbine blade vibration, vibration of aircraft wings, vibration of bridge suspension, which may cause a destruction of the bridge, such as the Tacoma Bridge, etc. In these systems, we are trying to eliminate these vibrations. 
The system was intended to measure a change of stress in elastic mechanical structures, construction and rock masses. The purpose of this study is to determine the possibilities of using this system for real objects such as bridges, dams, buildings, mines, etc. The sensitivity of this system, for small and large deformation, is higher than the sensitivity of other measurement systems (especially open systems). 
Tests on a single sample of sandstone were performed. They tried to examine the impact on the stress measurement parameters such as: position of sensors, position of actuator, and the influence of geometrical shape and dimensions of sample.

\begin{figure}[!ht]
\includegraphics[width=13cm]{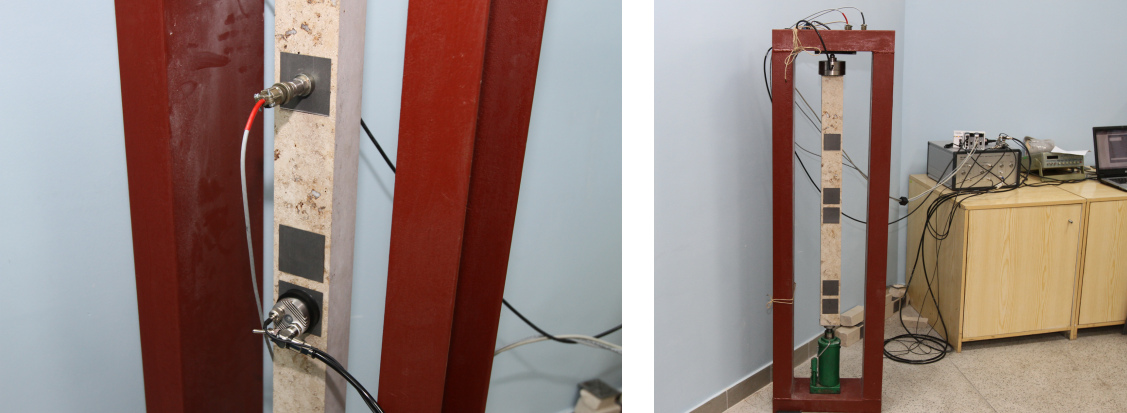}
\caption{The test stand for the endurance tests using the SAS System}
\label{fig:f7}
\end{figure}

The study used a sample of sandstone compressed in the frame by a hydraulic press (Fig. \ref{fig:f7}). In addition, the system was equipped with a force sensor to calculate the compressive stress in the beam in real time. Accelerations were measured by three accelerometers for three reasons:

- to determine what impact the distance from the emitter to the receivers has on the results and the position of resonances, \\
- to avoid a situation where the sensor is located in a resonance node, which would result in incorrect test results, \\
- to designate the velocity of wave propagation in a material with the correlation methods. \\

\begin{figure}[!ht]
\includegraphics[width=9cm]{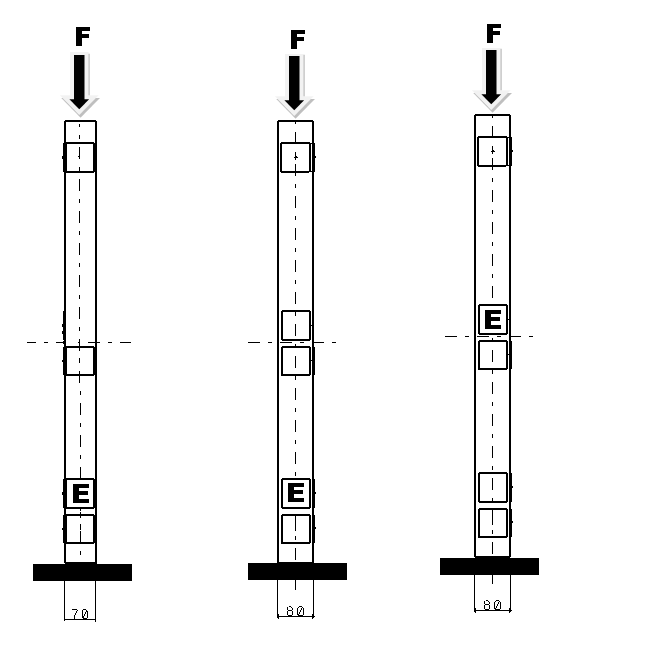}
\caption{Diagram of deployment of emitters on sandstone}
\label{fig:f8}
\end{figure}

The study was conducted in three configurations of the position of emitter (E). For a sandstone sample with rectangular cross-section of 60x70 emitting and receiving devices were mounted in the configuration shown in Figure \ref{fig:f8}. 

\section{Experimental  results}

\begin{figure}[!ht]
\includegraphics[width=13cm]{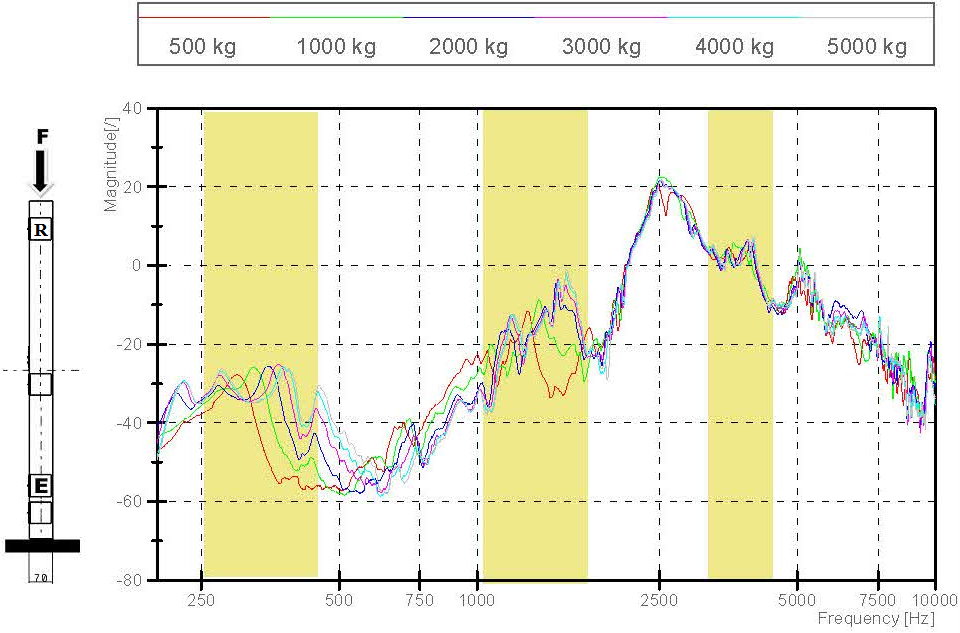}
\caption{Vibration transmissibility functions for different sample loads}
\label{fig:f9}
\end{figure}

\begin{figure}[!ht]
\includegraphics[width=11cm]{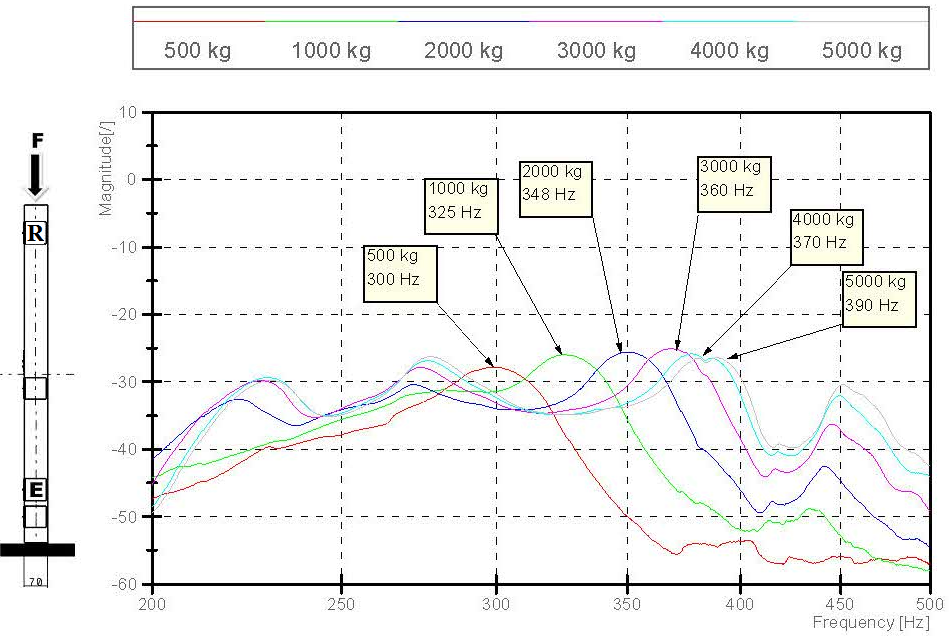}
\caption{Selected and expanded resonance peaks related to the change of stress}
\label{fig:f10}
\end{figure}

The SAS is designed for measurements of stress in elastic structures including rocks as well as mines, bridges, dams, buildings, etc. Therefore it is necessary to create a universal procedure to detect the change in the strain in objects mentioned above. The designers decided to use first the system to conduct tests in the open loop [3]. To the sample, charged with various compressive forces, was initially given the chirp signal emitted in the frequency range of 100 Hz to 20 kHz. This allowed to obtain the amplitude - frequency characteristics (Fig. \ref{fig:f9}) of the structure with visible resonances (Fig. \ref{fig:f10}).

\begin{figure}[!ht]
\includegraphics[width=11cm]{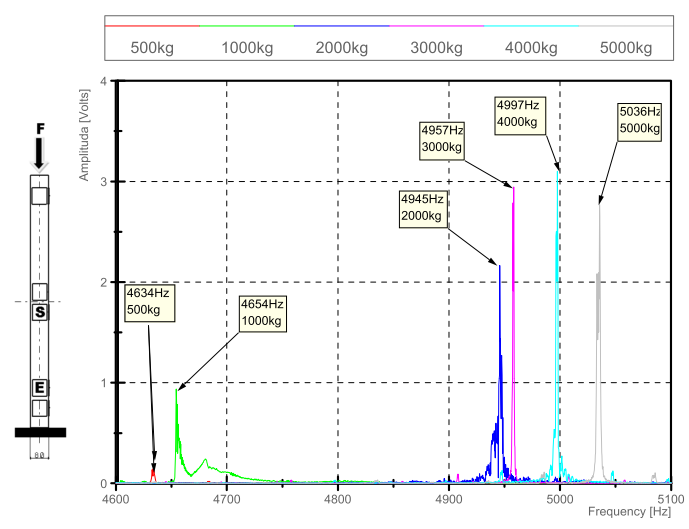}
\caption{Selected and expanded resonance peaks of closed loop system related to the change of stress}
\label{fig:f11}
\end{figure}

Figure \ref{fig:f10} presents the movement of the first resonance peak believed to be related to the change of stress in the sample. Changing the frequency respectively equals to 90 Hz. For the closed loop system, which is shown in figure \ref{fig:f11}, the frequency change at the same load change is as large as 450 Hz. In addition, the amplitude of vibration is 20-30 times greater than in an open loop system, which makes them more visible and easier to identify.

Thus our experiments have demonstrated an opportunity to detect stress changes in elastic geophysical objects: marble, sandstone and also concrete by means self-oscillating acoustic system (SAS). We hope that similar system might be helpful for stress change detection in rock mass and mines as well as civil engineering objects i.e. bridges, dams and buildings.

\subsection{Conclusions}

The main thesis presented in this paper is that self-exited acoustical system with the positive feedback allows monitoring of the change of the stresses in elastic structures. The usefulness of this phenomenon for different states of strain and the various sizes of test samples of sandstone, marble and concrete was examined to prove this property. This study confirmed the existence of the self-excitation phenomenon, which can be used to recognize the state of rock mass changeability.


\newpage
\begin{thebibliography}{99}
\bibitem{journal-1} J. Kwa\'{s}niewski, I. Dominik, J. Konieczny, K. Lalik, A. Sakeb, Application of self-oscillation system for stress measurement in sandstone bar. (MSM 2010: \textit{Mechatronic Systems and Materials}, 6th International Conference: 5–8 July 2010: Opole, Poland: eds. K. Kluger, E. Macha, R. Pawliczek, Opole University of Technology, 2010, ISBN 978-83-60691-78-6
\bibitem{journal-2} J. Kwa\'{s}niewski, I. Dominik, \textit{Autodyne effect in stress measurement in rocks, International Conference Source Science and Education}, : February 12–22, 2010, Colombo, Sri Lanka
\bibitem{journal-3} J. Kwa\'{s}niewski, I. Dominik, J. Konieczny, K. Lalik, Y. Kravtsov A.Sakeb, Experimental system for stress measurement in rocks, \textit{9th Conference on Active noise and vibration control methods} : Krak\'{o}w-Zakopane, Poland, May 24–27, 2009 , ISBN 83-89772-06-X
\bibitem{journal-4}  J. Kwa\'{s}niewski, I. Dominik, J. Konieczny, A.Sakeb, K. Lalik, Self-oscillating acoustical system application in industry in monograph : \textit{The improvement of the quality, reliability and long usage of technical systems and technological processes}, National Council of Ukraine for Mechanism and Machine Science, Federal State Unitary Enterprise Central, Institute of Aviation Motors (CIAM, Russia), Council of Scientific and Engineer, Union in Khmelnitsky Region, Khmelnitsky National University, 2010, ISBN 978-966-330-106-8
\bibitem{journal-5} T.G. Blaney, Infrared and Millimeter Waves, Vol 3, Part 2. NY: Acad. Press 1980 
\bibitem{journal-6} V.P. Shestopalov, Physical Foundations of the Millimeter and Submillimeter Waves Technique, Vol 2: Sources. Element Base. Radio Systems. Novel Scientific Trends. VSP, Zeist,   Netherlands, 1997.



\end{thebibliography}
\end{document}